\documentclass[]{aa}  
\usepackage{graphicx}
\usepackage{psfig}
\usepackage{txfonts}
\usepackage{amssymb}
\usepackage[below]{placeins}
\usepackage{natbib}
\bibpunct{(}{)}{;}{a}{}{,}
\begin{document}
   \title{Detection of hot gas in the filament connecting the clusters of galaxies Abell 222 and Abell 223}

   \subtitle{}

   \author{N. Werner\inst{1}
          \and
          A. Finoguenov\inst{2}
	  \and
	  J. S. Kaastra\inst{1,3}
	  \and
	  A. Simionescu\inst{2}
	  \and
	  J. P. Dietrich\inst{4}
	  \and
	  J. Vink\inst{3}
	  \and
	  H. B\"ohringer\inst{2}
	  }

   \institute{     SRON Netherlands Institute for Space Research, Sorbonnelaan 2, NL - 3584 CA Utrecht, the Netherlands \\ \email{n.werner@sron.nl}
         \and  Max-Planck-Institut f{\"u}r Extraterrestrische Physik, 85748 Garching, Germany
	 \and	Astronomical Institute, Utrecht University, P.O. Box 80000, NL - 3508 TA Utrecht, the Netherlands 
	 \and   ESO, Karl-Schwarzschild-Str. 2, 85748 Garching, Germany }

   \date{Received; accepted }

  \abstract
  % context heading (optional)
  % {} leave it empty if necessary  
   {About half of the baryons in the local Universe are invisible and - according to simulations - their dominant fraction resides in filaments connecting clusters of galaxies in the form of low density gas with temperatures in the range of $10^{5}<T<10^{7}$~K. The existence of this warm-hot intergalactic medium was never unambiguously proven observationally in X-rays. }
  % aims heading (mandatory)
   {We aim to probe the low gas densities expected in the large scale structure filaments by observing a filament connecting the massive clusters of galaxies A~222 and A~223 ($z=0.21$) which has a favorable orientation approximately along our line of sight. This filament has been previously detected using weak lensing data and as an over-density of colour selected galaxies.}
  % methods heading (mandatory)
   {We analyse X-ray images and spectra obtained in a deep observation (144~ks) of A~222/223 with XMM-Newton.}
  % results heading (mandatory)
   {  We present here observational evidence of the X-ray emission from the filament connecting the two clusters. We detect the filament in the wavelet decomposed soft band (0.5--2.0~keV) X-ray image with a $5\sigma$ significance. Following the emission down to $3\sigma$ significance, the observed filament is $\approx1.2$~Mpc wide. The temperature of the gas associated with the filament determined from the spectra is k$T=0.91\pm0.25$~keV and its emission measure corresponds to a baryon density of $(3.4\pm1.3)\times10^{-5}(l/15~\mathrm{Mpc})^{-1/2}$~cm$^{-3}$, where $l$ is the length of the filament along the line of sight. This density corresponds to a baryon over-density of $\rho/\left<\rho_{\mathrm{C}}\right> \approx150$. The properties of the gas in the filament are consistent with the results of simulations for the densest and hottest parts of the warm-hot intergalactic medium. }
  % conclusions heading (optional), leave it empty if necessary 
   {}


   \maketitle
%
%________________________________________________________________

\section{Introduction}

According to the standard theory of structure formation, the spatial distribution of matter in the Universe evolved from small perturbations in the primordial density field into a complex structure of sheets and filaments with clusters of galaxies at the intersections of this filamentary structure. The filaments have been identified in optical surveys of galaxies \citep[e.g.][]{joeveer1978,baugh2004,tegmark2004}, but the dominant fraction of their baryons is probably in the form of a low density warm-hot gas emitting predominantly soft X-rays. 

The existence of this warm-hot intergalactic medium (WHIM) permeating the cosmic-web is predicted by numerical hydrodynamic simulations \citep{cen1999,dave2001}, which show that in the present epoch ($z\lesssim1-2$) about 30\% to 40\% of the total baryonic matter resides in the filaments connecting clusters of galaxies. These simulations predict that the WHIM spans a temperature range of $10^{5}<T<10^{7}$~K and a density range of $4\times10^{-6}$~cm$^{-3}$ to $10^{-4}$~cm$^{-3}$,
which corresponds to $15$--$400$ times the mean baryonic density of the Universe. Gas at these temperatures and densities is difficult to detect \citep{dave2001}.

Several attempts have been made to detect the WHIM both in absorption and emission, but none of them succeeded in providing an unambiguous detection.
Marginal detection of \ion{O}{viii} and \ion{Ne}{ix} absorption features in XMM-Newton RGS spectra of distant quasars was reported from the vicinity of the Virgo and Coma clusters and was interpreted as indication of the presence of dense and hot WHIM concentrations near the clusters \citep{fujimoto2004,takei2007}.
A blind search for WHIM was performed by \citet{nicastro2005b}, who reported the discovery of two intervening systems at $z>0$ in absorption along the line of sight to the blazar Mrk~421 in the Chandra Low Energy Transmission Grating Spectrometer (LETGS) spectra. However, \citet{kaastra2006} carefully reanalysed the LETGS spectra, confirmed the apparent strength of several features, but showed that they are statistically not significant. Furthermore, the results reported by \citet{nicastro2005b} were neither confirmed by later Chandra LETGS observations nor by the XMM-Newton RGS data of Mrk~421 \citep{kaastra2006,rasmussen2006}. 
The detection of the WHIM in absorption with the current instruments will remain difficult, because the expected absorption features are weak and in the large searched redshift space to distant blazars there will always appear statistical fluctuations with a strength comparable to the strength of the absorption lines expected from the WHIM. A search for the WHIM from the densest and hottest parts of the large-scale structure filaments with the current instruments is more realistic in emission than in absorption. 

Searches in emission with ROSAT and XMM-Newton have found several candidates for WHIM filaments \citep{kull1999,zappacosta2002,finoguenov2003,kaastra2003}. However, the association of the detected soft emission with the WHIM is always difficult and uncertain. In some cases there is still a possible confusion with the Galactic foreground emission or with the Solar wind charge exchange emission that also produces \ion{O}{vii} lines and soft emission in excess to cluster emission. 

Since the X-ray emissivity scales with the square of the gas density, X-ray emission studies are most
sensitive to the densest gas concentrations of the WHIM which are, according to simulations, near clusters of galaxies. 
Therefore, the best approach to probe the physical properties of the WHIM is provided by those distant binary clusters between which the connecting bridge lies approximately along our line of sight. A filament between cluster pairs observed with such a favourable geometry has a higher surface brightness than filaments observed with a different orientation. By observing the variation of the surface brightness in X-ray images extracted in the low energy band, the presence of a WHIM filament can be revealed as a bridge between the clusters. This is a differential test with respect to the background and it is less sensitive to the systematic uncertainties. 

The most promising target for detecting the warm-hot gas in a filament is Abell~222/223, a close pair of massive clusters of galaxies at redshift $z \approx 0.21$ separated by $\sim$14$\arcmin$ in the sky
(which corresponds to a projected distance of $\sim2.8$~Mpc). 
Assuming that both clusters participate in the Hubble flow with no peculiar velocity, the observed redshift difference of $\Delta z=0.005\pm0.001$ \citep{dietrich2002} translates to a physical separation along the line of
sight of $15\pm3$~Mpc. \citet{dietrich2005} found indications for a filament connecting both clusters from weak lensing, optical, and X-ray (ROSAT) data. They found an X-ray bridge with a linear extent of about 1.3~Mpc (in projection) connecting the clusters at a $5\sigma$ significance level and an over-density of galaxies in the filament at a $7\sigma$ significance level. In the weak lensing image the filament was present at a $2\sigma$ significance.

Here we present the results of a new deep (144~ks) XMM-Newton observation of the bridge connecting the two clusters. We confirm its presence by both imaging and spectroscopy, and we estimate the temperature, density, entropy, and the total mass of the gas in the filament.
Throughout the paper we use $H_{0}=70$ km$\, $s$^{-1}\, $Mpc$^{-1}$, $\Omega_{M}=0.3$, $\Omega_{\Lambda}=0.7$, which imply a linear scale of 206~kpc\, arcmin$^{-1}$ at the redshift of $z=0.21$. 
Unless specified otherwise, all errors are at the 68\% confidence level for one interesting parameter ($\Delta \chi^{2}=1$).

\section{Observations and data analysis}

The pair of clusters Abell 222/223 was observed with XMM-Newton in two very nearby pointings centred on the filament on June 18 and 22, 2007 (revolutions 1378 and 1380) with a total exposure time of 144~ks. The calibrated event files were produced using the 7.1.0 version of the XMM-Newton Science Analysis System (SAS). After rejecting the time intervals with high particle background, we are left with 62.6~ks of good time for EPIC/MOS and with 34.4~ks for EPIC/pn. 

\subsection{Image analysis}

We coadded the final 0.5--2.0~keV and 2.0--7.5~keV images from all instruments and both pointings. The background was subtracted and the images were corrected for vignetting. The chip number 4 in EPIC/MOS1 was excluded from the analysis because of the anomalously high flux in the soft band \citep[for more details see][]{snowden2007}. 
In order to eliminate the flux pollution from the faint point sources due to the large wings of the point-spread function (PSF) of XMM-Newton, we applied an image restoration technique (Finoguenov et al. in prep) which uses a symmetric model for the PSF of XMM-Newton calibrated by \citet{ghizzardi2001}. The flux from the extended wings of the PSF of sources detected on small scales (sum of the 8\arcmin\arcmin\ and 16\arcmin\arcmin scale) is estimated performing a scale-wise wavelet analysis \citep{vikhlinin1998}. 
This estimated flux is used to subtract the PSF model prediction from the larger scales and to accordingly increase the flux on the smaller scales. The uncertainties in the point source subtraction are added to the total error budget in order to reduce the significance of the residuals.

\subsection{Spectral analysis}

\begin{table}
\begin{center}
\caption{The CXB components determined in a region free of clusters emission outside of the virial radii of the clusters. The unabsorbed fluxes are determined in the 0.3--10.0~keV band. $f$ indicates parameters kept fixed during spectral fitting.
\label{tab:back}}
\begin{tabular}{l|cccc}
\hline
\hline
Comp.			   &	k$T$/$\Gamma$	& Flux     		 			 \\
			   & (keV / phot. ind.) & (10$^{-12}$ erg cm$^{-2}$ s$^{-1}$ deg$^{-2}$) \\
\hline
LHB	 		   & k$T=0.08^f$  		& $3.4\pm1.1$ 	\\
GH 	 		   & k$T=0.17\pm0.03$      & 	$2.9\pm0.9$ 	\\  
EPL 			   & $\Gamma=1.41^f$    		& $22^f$       \\
PL			   & $\Gamma=0.78\pm0.16$	& $49\pm7$ \\
\hline
\end{tabular}
\end{center}
\vspace{-0.3cm}
\end{table}

We extracted the filament spectrum from a circular extraction region with a radius of 1.6\arcmin, located between the clusters, centred at a projected distance of 5.76\arcmin\ (1.19~Mpc) from both cluster cores. We determined the background spectrum using two different sets of regions which provide a consistency check. We used firstly a circular region with a radius of 3.2\arcmin\ located at a projected distance larger than 1.6~Mpc from both clusters, which is beyond their virial radii determined using weak lensing \citep[$r_{200}=1.28\pm0.11$~Mpc for A~222 and $r_{200}=1.55\pm0.15$~Mpc for A~223;][]{dietrich2005}. Secondly, we extracted two other circles with 1.6\arcmin\ radii located on the same central chip of the EPIC/MOS detectors, each of these circles being at the same projected distance to one of the clusters as the extraction region centred on the filament. Using the same chip and the same projected distance to the clusters for determining both the source and the background spectra ensures the best possible subtraction of both the instrumental and X-ray background component. Point sources were excluded from the spectral extraction regions. For the identification of point sources we also used archival Chandra data (45.7~ks). The regions used for spectral extraction, for both the filament and the background are shown in Fig.~\ref{filamentim}. 

We subtracted the instrumental background from all spectra using closed filter observations. Since the instrumental background level is changing from observation to observation and it is increasing with time over the years, we normalised the closed filter data for every extraction region by the 10--12~keV count rate for EPIC/MOS and by the 10--14~keV count rate for EPIC/pn. 
We then modeled the spectra in each background region set with the components described below and used these models to subtract the cosmic X-ray background from the filament spectrum.

We analysed the spectra using the SPEX package \citep{kaastra1996} in the 0.35 to 7~keV band for EPIC/MOS and 0.6--7~keV band for EPIC/pn, ignoring the 1.3--1.8~keV band because of the presence of strong instrumental lines. 
We fixed the Galactic absorption in our model to the value deduced from \ion{H}{i} radio data \citep[$N_{\mathrm{H}}=1.6\times10^{20}$~cm$^{-2}$,][]{kalberla2005}. We modelled the background with four components: with the extragalactic power-law (EPL) to account for the integrated emission of unresolved point sources \citep[assuming a photon index $\Gamma=1.41$ and a 0.3--10.0 keV flux of $2.2\times10^{-11}$~ erg s$^{-1}$ cm$^{-2}$ deg$^{-2}$ after extraction of point sources,][]{deluca2004}, with two thermal components to account for the local hot bubble (LHB) emission (k$T_1 = 0.08$~keV) and for the Galactic halo (GH) emission (k$T_2 \sim 0.2$~keV), and with an additional power-law to account for the contamination from the residual soft proton particle background. The best fitted fluxes, temperatures, and photon indices of the background components using the 3.2\arcmin\ circular extraction region are shown in Table~\ref{tab:back}.

We model the spectrum of the filament with a collisionally ionized plasma model (MEKAL), the metallicity of which is set to 0.2 Solar \citep[proto-Solar abundances by][]{lodders2003}, the lowest value found in the outskirts of clusters and groups \citep{fujita2007,buote2004b}. However, since the average metallicity of the WHIM is not known we report the results also for metallicities of 0.0, 0.1, and 0.3 Solar. The free parameters in the fitted model are the temperature and the emission measure of the plasma.

\section{Results}

\subsection{Imaging}

\begin{figure}[t]
\begin{center}
\includegraphics[width=\columnwidth,angle=0]{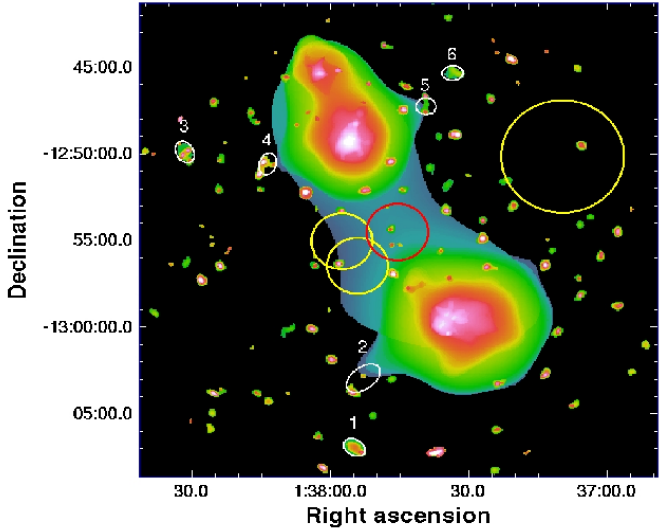}
\end{center}
\caption{Wavelet decomposed 0.5--2.0~keV image of Abell~222 (to the South) and Abell~223 (the two X-ray peaks to the North). We only show sources
with $>5\sigma$ detection, but for these sources we follow the emission down to $3\sigma$. The filament connecting the two massive clusters is clearly visible in the image. The regions used to extract the filament spectrum and the regions used to determine the background parameters are indicated with red and yellow circles, respectively.  }
\label{filamentim}
\vspace{-0.2cm}
\end{figure}

\begin{figure}[t]
\begin{center}
\includegraphics[width=0.75\columnwidth,angle=270]{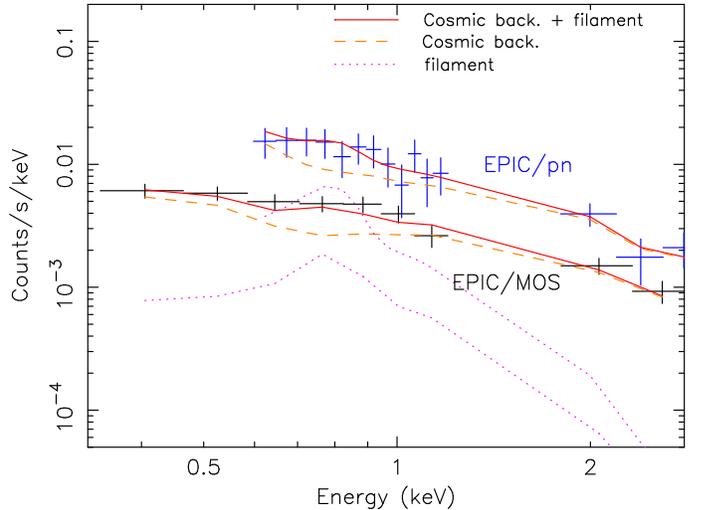}
\end{center}
\caption{The spectrum of the filament between the clusters A~222/223 - the data points on the top were obtained by EPIC/pn and below by EPIC/MOS. The contributions from the X-ray background and from the filament to the total model are shown separately. }
\label{spectrum}
\vspace{-0.2cm}
\end{figure}

We detect the filament in the wavelet decomposed soft band (0.5--2.0~keV) X-ray image with $5\sigma$ significance.
In Fig.~\ref{filamentim}, we show the wavelet decomposed image produced by setting the detection threshold to $5\sigma$ and following the emission down to $3\sigma$ \citep[for details of the wavelet decomposition technique see][]{vikhlinin1998}. There is a clear bridge in the soft emission between the clusters which originates from an extended component. We note that no such features are detected in any XMM-Newton mosaics of empty fields in observations with similar but also much larger depths. We do not detect the bridge between the clusters in the 2.0--7.5~keV image. 

To verify our result, we created an image modeling the X-ray emission from the two clusters without a filament with two beta models determined by fitting the surface brightness profile of A~222 and A~223. We applied on this image the wavelet decomposition algorithm, which did not reveal a bridge like the one observed between A~222 and A~223.

If the emission in the bridge between the clusters originated from a group of galaxies within the filament, then this group would be well resolved in the image. Groups of galaxies at the redshift of this cluster emit on spatial scales of $\approx$0.5\arcmin\ as shown in Fig.~\ref{filamentim} by the six extended sources detected at the $4\sigma$
significance around Abell~222/223 marked by ellipses numbered from 1 to 6. 
The filament seen in the image is about 6\arcmin\ wide, which at the redshift of the cluster corresponds to about 1.2~Mpc. 

The observed filament is by about an order of magnitude fainter than the ROSAT PSPC detected filament reported by \citet{dietrich2005}. 
The ROSAT detection was based on an image smoothed by a Gaussian with $\sigma=1.75$\arcmin\ and the filament could be the result of the combination of the emission from the cluster outskirts and from a few previously unresolved point sources between the clusters. 

The point sources (mostly AGNs) detected by XMM-Newton and Chandra do not show an overdensity between the clusters and the uniform contribution from the distant unresolved AGNs was subtracted from our image.
Our detection limit for point sources is $\sim0.8\times10^{-15}$~erg~s$^{-1}$~cm$^{-2}$, which corresponds to a luminosity of $10^{41}$~erg~s$^{-1}$ for the faintest resolved galaxy at the cluster redshift. The number of X-ray emitting galaxies with this luminosity required to account for the observed emission in the filament would be $\sim10$ times larger than is expected from the distribution function of \citet{hasinger1998}, assuming a filament overdensity of $\rho/\left<\rho\right>\sim100$.

\subsection{Spectroscopy}

The clusters of galaxies Abell 222 and Abell 223 are bright and relatively hot. The temperature of A~222, extracted from a region with a radius of 2\arcmin\ is $4.43\pm0.11$~keV and the temperature of southern core of A~223 extracted within the same radius is $5.31\pm0.10$~keV. However, in this paper we focus on the bridge connecting the two clusters. 

As shown in the Fig.~\ref{spectrum} the spectrum extracted from the filament shows between $\sim0.5$ and $\sim1.0$~keV $\sim30$\% emission in excess to the background model (determined outside of the cluster virial radii, see Table~1). We fit this excess emission with a thermal model with a temperature of k$T=0.91\pm0.24\pm0.07$~keV, where the first error is the statistical error and the second error is due to the uncertainty on the background level, determined by varying the background components in the spectral fit within their $1\sigma$ errors. 
The best fit emission measure ($EM=\int n_{\mathrm{H}}\, n_{\mathrm{e}} \mathrm{d}V$, where $n_\mathrm{H}$ and $n_\mathrm{e}$ are the proton and electron number densities and $V$ is the volume of the emitting region) of the filament is: $(1.72\pm0.50\pm0.45)\times10^{65}$~cm$^{-3}$, where the first error is again the statistical error and the second error is due to the uncertainty on the background level. Assuming both errors have a Gaussian distribution, the emission measure of the plasma in the filament within our extraction region is $EM=(1.72\pm0.67)\times10^{65}$~cm$^{-3}$ (2.6$\sigma$). Fitting individually EPIC/MOS and EPIC/pn the best fit temperature is $0.94\pm0.23$~keV and $0.85\pm0.30$~keV, respectivelly, and the statistical significance of the excess emission is 3.5$\sigma$ and 1.8$\sigma$. The derived emission measure depends on the assumed metallicity. For a higher metallicity we obtain a lower emission measure (for $Z=0.3$ Solar, $EM=1.3\times10^{65}$~cm$^{-3}$), while for a lower metallicity the obtained emission measure will be higher (for $Z=0.1$~Solar, $EM=2.4\times10^{65}$~cm$^{-3}$ and for $Z=0$~Solar, $EM=4.13\times10^{65}$~cm$^{-3}$). The plasma temperature does not change significantly as a function of the assumed metallicity. The 0.3--10.0 keV luminosity of the filament within our extraction region is $1.4\times10^{42}$~erg~s$^{-1}$. In the COSMOS survey, a number of groups have been detected at these luminosities and redshifts, but their typical spatial extent ($r_{500}\sim1.3$\arcmin, Finoguenov et al. in prep.) is much smaller than that of this filament.

By determining the background properties using the two extraction regions at the central chip of EPIC/MOS and fitting only the EPIC/MOS data, we obtain a best-fit emission measure and temperature for the filament of $EM=(1.7\pm0.6)\times10^{65}$~cm$^{-3}$ and k$T=0.80\pm0.19$~keV. These values are consistent with those obtained using a background region located at a larger distance on a different part of the detector. This minimizes chip-to-chip variations of the instrumental noise and allows to include some residual cluster emission in the form of an increased best-fit flux of the power-law model component. However, part of this increased power-law flux could also be due to emission from the filament in the background regions, resulting in the oversubtraction of the filament.

\section{Discussion}

We detect X-ray emission from a bridge connecting the clusters A~222 and A~223. The temperature of the gas associated with this bridge is k$T=0.91\pm0.25$~keV. If this gas is the intra-cluster medium (ICM) at the cluster outskirts, then assuming an emitting volume with a line of sight depth of 2.5~Mpc, our best fit emission measure corresponds to a density of $\sim1\times10^{-4}$~cm$^{-3}$ and the entropy of the gas is  $s=kT/n^{2/3}\sim420$~keV~cm$^{2}$. Because of the heating by accretion shocks, the cluster entropies rise towards the outskirts to values higher than 1000~keV~cm$^{2}$ outside 0.5R$_{200}$ \citep{pratt2006a} and the entropy is expected to rise further out to the virial radius. The low entropy of the bridge implies that the emitting gas is in a filament which was not yet reached by the shock-heating operating on the ICM in the cluster outskirts.

If the redshift difference between the clusters is only due to the Hubble flow, their line of sight distance difference is $15\pm3$~Mpc. Assuming that the observed emission originates from hot gas within the filament connecting the clusters the mean baryon density of the gas is $3.4\times10^{-5}(l/15~\mathrm{Mpc})^{-1/2}$~cm$^{-3}$ (where $l$ is the length of the filament along the line of sight), which corresponds to 150 times the mean baryon density of the Universe. From this density (assuming $l=15$~Mpc) and the temperature of k$T=0.91$~keV we obtain an entropy $s=kT/n^{2/3}\sim870$~keV~cm$^{2}$. For a filament with a cylindrical shape, which is 1.2~Mpc wide as indicated by the images, assuming a constant density we obtain a baryon mass of $1.8\times10^{13}\, (l/15~\mathrm{Mpc})^{1/2}$~M$_\odot$. 
Assuming a baryon fraction of 0.16, the total mass of the filament would be $\approx1.1\times10^{14}$~M$_\odot$. 

The inferred properties of the gas in the filament are remarkably consistent with the results of the simulations by \citet{dolag2006}. They predict that within a distance of $\approx15$~Mpc from a massive cluster, filaments connecting clusters have a baryon over-density of $\rho/\left<\rho_{\mathrm{C}}\right> > 100$ inside a radius of $\approx$1~Mpc from the centre of the filament. They also found that the temperature of the gas along the bridges connecting some nearby clusters can be around $10^7$~K, which is the temperature of the gas that we see between A~222 and A~223. The predicted X-ray surface brightness of filaments is according to \citet{dolag2006} at most $10^{-16}$~erg~s$^{-1}$~cm$^{-2}$~arcmin$^{-2}$. However, this predicted surface brightness was derived for zero metallicity (for a metallicity of $Z=0.2$ Solar the predicted X-ray flux would increase due to line emission by a factor of $\sim1.6$) and for the case of looking at the filament edge on. Line emission and a favourable geometry can increase the surface brightness to our observed value of $10^{-15}$~erg~s$^{-1}$~cm$^{-2}$~arcmin$^{-2}$.

If the observed redshift difference is entirely due to peculiar velocities ($\Delta z=0.005$ corresponds to $\Delta v_{\mathrm{r}}=1500$~km~s$^{-1}$) and both clusters are actually at the same distance from us, then the virial radii of the clusters partially overlap and they are at the onset of a merger. 
However, if this is the case, then we do not expect to observe such a large amount of gas at these low temperatures between the clusters. Most of the mass from a previously present filament would be accreted by the clusters and the gas in the interaction region would be compressed and shock-heated to temperatures significantly higher than those observed. The redshift difference thus cannot be entirely due to the peculiar velocities. 

The temperature and the average density of the observed gas associated with the filament indicates that we are detecting the hottest and densest phase of the WHIM. It is only detectable because of the favourable geometry of the massive filament connecting two large clusters of galaxies. However, according to the simulations \citep[e.g.][]{dave2001} the dominant fraction of the WHIM resides in a lower temperature and density phase, the existence of which still remains to be proven observationally. The detection of the dominant fraction of the WHIM will only be possible with dedicated future instrumentation \citep[e.g. see ][]{paerels2008}.

In order to further investigate the physical properties of the densest and hottest gas phase permeating the cosmic web, the observed sample of cluster pairs where the large scale structure filaments can be observed along our line of sight needs to be enlarged by other promising systems.

\begin{acknowledgements}
AF acknowledges support from BMBF/DLR under grant 50 OR 0207 and MPG.
This work is based on observations obtained with XMM-Newton, an ESA science mission with instruments
and contributions directly funded by ESA member states and the USA (NASA). The Netherlands Institute
for Space Research (SRON) is supported financially by NWO, the Netherlands Organization for Scientific Research. 
\end{acknowledgements}

\bibliographystyle{aa}
\bibliography{clusters}

\end{document}